\documentclass[a4paper,12pt]{article}
\usepackage{amssymb}
\usepackage{amsmath, amsthm, amsfonts, amssymb, graphicx}
\usepackage{subfigure}
\usepackage{indentfirst}
\usepackage{bm}
\usepackage{indentfirst}
\usepackage{picins}
\usepackage{abstract}
\usepackage{dsfont}
\allowdisplaybreaks

\newtheorem{defi}{Definition}

\newtheorem{theo}{Theorem}
\newtheorem{lemm}[theo]{Lemma}
\newtheorem{coro}[theo]{Corollary}
\newtheorem{prop}[theo]{Proposition}

\newtheorem*{rem}{Remark}
\begin{document}
\normalsize

\title{{\bf Fast Algebraic Attacks and Decomposition of
Symmetric Boolean Functions}
\thanks{This work was supported by the National Natural Science Foundation of China under Grants 10971246 and 60970152.}}

\author{Meicheng Liu$^{\dag~\ddag}$ and Dongdai Lin$^{\dag}$\\
\normalsize
$^{\dag}$ The State Key Laboratory of Information Security,
Institute of Software\\
\normalsize
Chinese Academy of Sciences, Beijing 100190, China\\
\normalsize
$^{\ddag}$ Graduate University of Chinese Academy of
Sciences, Beijing 100049, China\\
\normalsize
E-mail: meicheng.liu@gmail.com, ddlin@is.iscas.ac.cn
}
\date{}
\maketitle
\setlength{\oddsidemargin}{0cm}  
\setlength{\evensidemargin}{\oddsidemargin}
\setlength{\textwidth}{13.5cm} \vspace{1pt}

\normalsize
\begin{center}
\parbox{\textwidth}{{\textbf{Abstract}\quad Algebraic and fast algebraic attacks are power tools to analyze stream ciphers.
A class of symmetric Boolean functions with maximum algebraic immunity were found vulnerable to fast algebraic attacks at EUROCRYPT'06.
Recently, the notion of $\mathcal{AAR}$ (algebraic attack
resistant) functions was introduced as a unified measure of protection
against both classical algebraic and fast algebraic attacks. In this correspondence, we first give a decomposition of symmetric Boolean
functions, then we show that almost all symmetric Boolean functions, including these functions
with good algebraic immunity, behave badly against fast algebraic
attacks, and we also prove that no symmetric Boolean
functions are $\mathcal{AAR}$ functions.
Besides, we improve the relations between algebraic degree and algebraic
immunity of symmetric Boolean functions.
}}
\end{center}

\section{Introduction}
{B}{oolean} functions are frequently used in the design of stream ciphers, block ciphers and hash functions. One of the most vital roles in cryptography of Boolean functions is to be used as filter and combination generators of stream ciphers based on linear feedback shift registers (LFSRs). Symmetric functions
are an interesting subclass of Boolean functions for their advantage in both implementation complexity and storage space (see \cite{Canteaut}).

In recent years, {a}{lgebraic} and fast algebraic attacks
\cite{Courtois1,Meier,Courtois2,Armknecht04} have been regarded as a great threat against LFSR-based stream ciphers. 
These attacks use cleverly over-defined
systems of multi-variable nonlinear equations to recover the secret
key. Algebraic attacks lower the degree of the equations
by multiplying a nonzero function while fast algebraic attacks by linear combination.
Thus algebraic immunity (AI) was introduced in \cite{Meier} to measure the ability of
Boolean functions to resist algebraic attacks while the notion of $\mathcal{AAR}$ (algebraic attack
resistant) functions in \cite{Pasalic} as a unified measure of protection
against both classical algebraic and fast algebraic attacks.

The maximum algebraic immunity (MAI) of $n$-variable Boolean
functions is $\lceil\frac{n}{2}\rceil$ \cite{Courtois1}.
The majority function 
achieves MAI \cite{Dalai,Braeken}.
For odd $n$, the majority function is the only
symmetric MAI functions, up to addition of a constant \cite{sys_li,sys_qu}. However, the majority function was found
vulnerable to fast algebraic attacks in \cite{Armknecht} at EUROCRYPT'06.

All the symmetric MAI functions on $2^m$ variables
were obtained in \cite{sys_qu2,sys_liu} and were proven
having algebraic degree $2^{m-1}$ or $2^m$. Moreover, all the symmetric functions on $2^{m}+1$ variables with sub-MAI $2^{m-1}$
were derived in \cite{{sys_liao}}. A general method to construct symmetric MAI functions was further provided in \cite{sys_qu3}. Nevertheless, we find all these functions but few very vulnerable to fast algebraic attacks despite their resistance against classical algebraic attacks.

A preprocessing of fast algebraic attacks on LFSR-based stream ciphers, which use a Boolean function $f$ as the filter or combination generator, is to find a function $g$ of small degree
such that the multiple $gf$ has degree not too large. For
any pair of integers $(e,d)$
such that $e+d\geq n$, there is a nonzero function $g$ of degree at
most $e$ such that $gf$ has degree at most $d$ \cite{Courtois2}.
We concentrate on the minimum of $e+d$ and
introduce the notion of fast algebraic immunity (FAI), which generalizes the notion of $\mathcal{AAR}$.
The full fast algebraic immunity (FFAI) of Boolean functions is $n$ and FFAI functions are equivalent to  $\mathcal{AAR}$ functions.
In this correspondence, the fast algebraic immunity of symmetric Boolean functions is studied.
It's found that any symmetric function is a composition of a Boolean function and elementary symmetric functions with degree equal to a power of 2 and that the set of all symmetric functions with degree at most $2^k-1$ is a ring generated by $\sigma_1,\sigma_2,\sigma_4,\cdots,\sigma_{2^{k-1}}$ and isomorphic to $\mathbf{B}_k$. It's further shown that almost all symmetric Boolean functions behave badly against
fast algebraic attacks. For one thing, any symmetric function with degree not equal to a power of 2 has FAI strictly less than its degree, for another, in the case $n$ close to
$2^{\lfloor\log_2n\rfloor}$, symmetric functions with AI at least $2^{\lfloor\log_2n\rfloor}/2$ 
have FAI close to ${n}/{2}$, which is almost the worst case against fast algebraic attacks. 
Unfortunately, all but few symmetric functions shown to be immune to classical algebraic attacks in the previous literatures, such as \cite{sys_qu2,sys_liu,sys_liao,sys_qu3}, fall into the case that $n$ is either equal to or a little more than $2^{\lfloor\log_2n\rfloor}$.
One (or more) function $g$ with small degree, such that $gf$ has degree not
large, is straightway derived from the SANFV of $f$, while the algorithm proposed in \cite{Armknecht} at EUROCRYPT'06 to determine $g$ and $gf$ for a symmetric function $f$ has
complexity $\mathcal{O}(n^3)$.
Furthermore, it's proven that there exist no symmetric FFAI (i.e. $\mathcal{AAR}$) functions.
%
Lastly, the relations between algebraic
degree and algebraic immunity of symmetric Boolean functions are improved.
$2^{\lfloor\log_2(2a-1)\rfloor}$ is the lower degree of symmetric
functions with AI $a$. This bound is tight for symmetric MAI functions.

The remainder of this correspondence is organized as follows. In Section \ref{sec:Pre}, some basic concepts are provided and the notion of fast algebraic immunity is introduced, while Section \ref{sec:Decom} presents the decomposition of symmetric functions.
Section \ref{sec:FAA} studies the fast algebraic immunity of symmetric functions and Section \ref{sec:Relations} discusses the relations between algebraic
degree and algebraic immunity of symmetric functions. Section \ref{sec:Con} concludes the correspondence.

\section{Preliminary}\label{sec:Pre}
{A}{n} $n$-variable Boolean function is a mapping from $\mathbb{F}_2^n$ into $\mathbb{F}_2$, where $\mathbb{F}_2$ denote the binary field.
A Boolean function is said to
be symmetric if its output is invariant under any permutation of its
input bits.
Denote by $\mathbf{B}_n$ (resp. $\mathbf{SB}_n $) the set of all Boolean functions (resp. symmetric Boolean functions) on $n$ variables.
Any function $f\in\mathbf{B}_n$ can be uniquely represented as a truth table
$$
 f=[f(0, 0, \cdots, 0), f(1, 0, \cdots, 0),\cdots, f(1, 1, \cdots, 1)]\in\mathbb{F}_2^{2^n},
$$
or as a multivariate polynomial over $\mathbb{F}_2$, called the algebraic
normal form (ANF),
$$
 f(x)=\sum_{c=(c_1,c_2,\cdots,c_n)\in\mathbb{F}_2^n}a_{c}x_1^{c_1}x_2^{c_2}\cdots
x_n^{c_n},~a_c\in \mathbb{F}_2.
$$
The algebraic degree of $f$, denoted by $deg(f)$, is given by
$\max_{a_c\neq 0}wt(c)$, where $wt(c)$ 
denote the Hamming weight of $c$.
Any function $f\in
\mathbf{SB}_n $  can be uniquely represented as a vector
$$v_f= (v_f
(0), v_f (1),\cdots, v_f (n))\in\mathbb{F}_2^{n+1},$$ where $v_f (i)$
represents the function value for vectors of weight $i$. Let
$\sigma_i$ be the $i$-th elementary symmetric function of $x_1,
x_2,\cdots, x_n.$ The symmetric function $f$ can also be uniquely represented
as $$f(x)=\sum_{i=0}^{n}\lambda_f (i)\sigma_i,~\lambda_f (i)\in
\mathbb{F}_2.$$ The vector $\lambda_f= (\lambda_f (0),\lambda_f
(1),\cdots,\lambda_f (n))$ is called the simplified algebraic normal
form vector (SANFV) of $f$.
 More properties of symmetric Boolean functions can be found in
\cite{Canteaut}.

The algebraic immunity of Boolean functions is defined as follows.
\begin{defi} \cite{Meier}
Let $f$ be an $n$-variable Boolean function. The algebraic immunity (AI)
of $f$, denoted by  $\mathcal{AI}(f)$, is defined as
$$\mathcal{AI}(f)=\min_{g\neq 0}\{\deg(g)|gf=0\text{~or~}g(f+1)=0\}.$$
\end{defi}

To resist fast algebraic attacks, the Boolean function $f$ shouldn't admit a function $g$ of small degree
such that the multiple $gf$ has degree not too large.
There are several notions of the immunity of Boolean functions against fast algebraic attacks in previous literatures, such as \cite{Gong}, but they separately treat the two parameters $\deg(g)$ and $\deg(gf)$.
Recently, the notion of $\mathcal{AAR}$ (algebraic attack
resistant) functions was introduced in \cite{Pasalic} as a unified measure of protection
against classical algebraic attacks as well as fast algebraic attacks.
\begin{defi}\label{AAR}\cite{Pasalic}
Let $f$ be an $n$-variable Boolean function. The function $f$ is called $\mathcal{AAR}$ if $f$ has MAI and
$\deg(g)+\deg(gf)\geq n$ for any function $g$ with $1\leq\deg(g)<{n}/{2}$.
\end{defi}
However, $\mathcal{AAR}$ is too restrictive to achieve. A 9-variable $\mathcal{AAR}$ function was observed in \cite{CarlFeng}.
A class of almost $\mathcal{AAR}$ functions were constructed in \cite{Pasalic} by iteration.
While it's still unknown whether there are $\mathcal{AAR}$ functions for any $n$.

For any Boolean function $f$, from the definition of AI there always exists a function $g$ of degree equal to $\mathcal{AI}(f)$ such that $gf=0$ or $gf=g$.
Therefore the minimum of $\deg(g)+\deg(gf)$ is smaller than or equal to $2\mathcal{AI}(f)$.
The notion of fast algebraic immunity is introduced as follows.
\begin{defi}\label{FAI}
Let $f$ be an $n$-variable Boolean function. The fast algebraic
immunity (FAI) of the function $f$, denoted by  $\mathcal{FAI}(f)$,
is defined as
$$\mathcal{FAI}(f)=\min_{g:1\leq\deg(g)<\mathcal{AI}(f)
}\{2\mathcal{AI}(f), \deg(g)+\deg(gf)\}.$$
\end{defi}
From the above definition, we know that $\mathcal{AI}(f)+1\leq\mathcal{FAI}(f)\leq \deg(f)+2$
since $\mathcal{AI}(f)+1\leq\deg(g)+\deg(gf)$ for any nonconstant function $g$ with degree less than $\mathcal{AI}(f)$ and $\deg(l)+\deg(lf)\leq\deg(f)+2$ for any affine function $l$.

For
any pair of integers $(e,d)$
such that $e+d\geq n$, there is a nonzero function $g$ of degree at
most $e$ such that $gf$ has degree at most $d$ \cite{Courtois2}.
Hence, the full fast algebraic
immunity (FFAI) of $n$-variable Boolean functions is $n$. It is clear that any Boolean function has AI greater than or equal to a half of its FAI.
Therefore FFAI functions are also MAI functions and are equivalent to $\mathcal{AAR}$ functions.
By almost FFAI or almost $\mathcal{AAR}$ functions we mean Boolean functions with FAI $n-1$.

\section{Decomposition of symmetric Boolean functions}\label{sec:Decom}
Thereinafter, ${k\choose i}$ may be regarded as ${k\choose i}~mod~2\in\mathbb{F}_2$ if there is no ambiguousness.
\begin{lemm}\label{basic}
Let $\sigma_i$ and $\sigma_j$ be $i$-th and $j$-th elementary symmetric Boolean
functions on $n$ variables, $0\leq i, j\leq n$. Then we have
$$\sigma_i\sigma_j=\sum_{k=j}^{j+i}{k\choose i}{i\choose k-j}\sigma_k
=\sum_{k=0}^{n}{k\choose i}{i\choose k-j}\sigma_k.$$
In particular, $\sigma_j^2=\sigma_j.$
\end{lemm}
\begin{proof}
Expanding the product $\sigma_i\sigma_j$
gives ${n\choose i}{n-i\choose k-i}{i\choose k-j}$ monomials 
with degree $k$ for $j\leq k \leq j+i$ and $0\leq k\leq n$.
Since the product $\sigma_i\sigma_j$ is also a symmetric
function and $\sigma_k$ consists
of ${n\choose k}$ monomials with degree $k$, the coefficient of
$\sigma_k$ in $\sigma_i\sigma_j$ equals to ${n\choose i}{n-i\choose k-i}{i\choose
k-j}/{n\choose k}={k\choose i}{i\choose k-j}$.
Since $f^2=f$ for any Boolean function $f$, we also have $\sigma_j^2=\sigma_j$ .
\end{proof}


\begin{coro}\label{basic_coro:1}
\begin{enumerate}
  \item $\sigma_{2^{s_1}+\cdots+2^{s_k}}=\sigma_{2^{s_1}}\sigma_{2^{s_2}}\cdots\sigma_{2^{s_k}}$ for pairwise different $s_1,s_2,\cdots,s_k$.
  \item If $t\geq 1$ and $j<2^s$ then $\sigma_{t\cdot2^s+j}=\sigma_{t\cdot2^s}\sigma_j$.
\end{enumerate}
\end{coro}
\begin{proof}
1)
Consider the polynomial
$(1+x)^{2^s}\in\mathbb{F}_2[x]$. Since
$1+x^{2^s}=(1+x)^{2^s}=\sum_{k=0}^{2^s}{2^s\choose k}x^k$, we have
${2^s\choose k}=1$ if and only if $k=0,2^s$. By Lemma \ref{basic} we have
\begin{equation}\label{eq:1}
\sigma_{2^s}\sigma_{j}=\sum_{k=j}^{2^s+j}{k\choose 2^s}{2^s\choose k-j}\sigma_k={j\choose 2^s}\sigma_{j}+{2^s+j\choose
2^s}\sigma_{2^s+j}.
\end{equation}
If $j<2^s$, then
${j\choose 2^s}=0$ and ${2^s+j\choose 2^s}=1$ (considering the
polynomial $(1+x)^{2^s+j}\in\mathbb{F}_2[x]$), and by Eq.(\ref{eq:1}) we have
$\sigma_{2^s}\sigma_{j}=\sigma_{2^s+j}$.
Assuming $s_1<s_2<\cdots<s_k$ without loss of generality, since $2^{s_{i-1}}+\cdots+2^{s_{1}}<2^{s_i}$ for $2\leq i\leq k$, we have  $\sigma_{2^{s_i}}\sigma_{2^{s_{i-1}}+\cdots+2^{s_{1}}}=\sigma_{2^{s_i}+\cdots+2^{s_{1}}}$ and therefore $\sigma_{2^{s_k}+\cdots+2^{s_1}}
=\sigma_{2^{s_k}}\sigma_{2^{s_{k-1}}+\cdots+2^{s_1}}=\cdots=\sigma_{2^{s_k}}\sigma_{2^{s_{k-1}}}\cdots\sigma_{2^{s_1}}.$

2) Let $t\cdot 2^s=2^{s}+2^{s_1}+\cdots+2^{s_k}$, $s<s_1<\cdots<s_k$. From 1) we know $\sigma_{t\cdot 2^s}=\sigma_{2^{s_k}}\cdots\sigma_{2^{s_1}}\sigma_{2^s}$, and hence $\sigma_{t\cdot 2^{s}}{\sigma_j}=\sigma_{2^{s_k}}\cdots\sigma_{2^{s_1}}\sigma_{2^s}{\sigma_j}
=\sigma_{2^{s_k}}\cdots\sigma_{2^{s_1}}\sigma_{2^s+j}=\cdots=\sigma_{2^{s_k}+\cdots+2^s+j}=\sigma_{t\cdot2^s+j}.$
\end{proof}

From the above corollary we obtain the following results proven
in \cite{Braeken}.

\begin{coro}\label{basic_coro:2}
Let $m=\lfloor\log_2n\rfloor$ and $j=\sum_{k=0}^{m}j_k2^{k}$, $j_k\in\{0,1\}$.
Then we have
\begin{enumerate}
  \item $\sigma_j=\sigma_{1}^{j_0}\sigma_{2}^{j_{1}}\sigma_{4}^{j_{2}}\cdots\sigma_{2^m}^{j_m}$.
  \item If $n-2^m<j<2^m$ then $\sigma_{2^m}\sigma_{j}=0$.
  \item $\sigma_i\sigma_j=\sigma_{i\vee j}$ where $\vee$ means OR operation.
\end{enumerate}
\end{coro}
\begin{proof}
1) It is affirmed by Corollary \ref{basic_coro:1} (1).

2) It is derived from the fact that $\sigma_{2^m}\sigma_{j}=\sigma_{2^m+j}$ for $j<2^m$ by Corollary \ref{basic_coro:1} (2) and there is no $\sigma_{2^m+j}$ for $2^m+j>n$.

3) It can be deduced from 1) since $\sigma_{2^s}^2=\sigma_{2^s}.$
\end{proof}


Thanks to the work above, the decomposition of symmetric Boolean functions is given as follows.
\begin{theo}[Decomposition of symmetric Boolean functions]\label{factor}
Let $f\in\mathbf{SB}_n$ and $m=\lfloor \log_2n\rfloor$.
\begin{enumerate}
  \item The symmetric function $f$ is a composition of an $(m+1)$-variable Boolean function $F_{m+1}$ and elementary symmetric functions $\sigma_1,\sigma_2,\sigma_4,\cdots,\sigma_{2^{m}}$:
$$f(x)=F_{m+1}(\sigma_1,\sigma_2,\sigma_4,\cdots,\sigma_{2^{m}}).$$
In particular, if $f$ has degree at most $2^k-1$, then
$f(x)=F_k(\sigma_1,\sigma_2,\sigma_4,\cdots,\sigma_{2^{k-1}})$, $F_k\in\mathbf{B}_k$.
  \item Furthermore we have \begin{align*}
f(x)=\sum_{i=k}^{m}
\sigma_{2^i}f_{i}(x)+f_{k}^-(x),
\end{align*}
where $f_i$ ($k\leq i\leq m$) and $f_{k}^-$ are symmetric functions
of degree at most $2^i-1$ and $2^k-1$.
\end{enumerate}
\end{theo}
\begin{proof}
1) Let $f(x)=\sum_{j=0}^{n}\lambda_f(j)\sigma_j,~\lambda_f(j)\in
\mathbb{F}_2$. By Corollary \ref{basic_coro:2}, we have $\sigma_j=\sigma_{1}^{j_0}\sigma_{2}^{j_{1}}\sigma_{4}^{j_{2}}\cdots\sigma_{2^m}^{j_m}$ and hence
$$f(x)=\sum_{j=0}^{n}\lambda_f(j)\sigma_j
=\sum_{j=0}^{n}\lambda_f(j)\sigma_{1}^{j_0}\sigma_{2}^{j_{1}}\sigma_{4}^{j_{2}}\cdots\sigma_{2^m}^{j_m}.$$
Let $F_{m+1}\in \mathbf{B}_{m+1}$ and $$F_{m+1}(y_1,y_2,\cdots,y_{m+1})=\sum_{0\leq j\leq n}\lambda_f(j)y_1^{j_0}y_2^{j_1}\cdots y_{m+1}^{j_m}.$$
Then $$f(x)=F_{m+1}(\sigma_1,\sigma_2,\sigma_4,\cdots,\sigma_{2^{m}}).$$
The same proof shows $f(x)=F_k(\sigma_1,\sigma_2,\sigma_4,\cdots,\sigma_{2^{k-1}})$ with $F_k\in\mathbf{B}_k$ when $\deg(f)\leq 2^k-1$.

2) 
Since $F_{m+1}$ is a Boolean function, we can write $F_{m+1}(y_1,\cdots,y_{m+1})=\sum_{i=k}^{m}y_{i+1}F_{i}(y_1,\cdots,y_{i})+F_{k}^-(y_1,\cdots,y_{k}), k\geq 1.$
Therefore
\begin{align*}
f(x)=&F_{m+1}(\sigma_1,\sigma_2,\sigma_4,\cdots,\sigma_{2^{m}})\\
=&\sum_{i=k}^{m}\sigma_{2^i}F_{i}(\sigma_1,\sigma_2,\sigma_4,\cdots,\sigma_{2^{i-1}})\\
&+F_{k}^-(\sigma_1,\sigma_2,\sigma_4,\cdots,\sigma_{2^{k-1}}).
\end{align*}
%
%
%
%
Let $f_i(x)=F_i(\sigma_1,\sigma_2,\sigma_4,\cdots,\sigma_{2^{i-1}})$ ($k\leq i\leq m$) and $f_{k}^-(x)=F_k^-(\sigma_1,\sigma_2,\sigma_4,\cdots,\sigma_{2^{k-1}})$. The symmetric function $f_i$ has degree at most $2^i-1$ since the degree of $\sigma_{1}^{j_0}\sigma_{2}^{j_{1}}\sigma_{4}^{j_{2}}\cdots\sigma_{2^{i-1}}^{j_{i-1}}$ can not exceed $1+2+4+\cdots+2^{i-1}=2^i-1.$
Similarly, $f_k^-$ has degree at most $2^k-1$.
\end{proof}
%
%

Note that $\lceil
\log_2(n+1)\rceil=\lfloor \log_2n\rfloor+1$. Theorem \ref{factor}
shows that an
$n$-variable symmetric Boolean function corresponds to a $\lceil
\log_2(n+1)\rceil$-variable Boolean function. Furthermore, a symmetric Boolean function of degree at most $2^k-1$
corresponds to a $k$-variable Boolean function.

\begin{theo}\label{basic_coro:Ring}
Let $1\leq 2^k\leq n+1$. Then the set of all functions in $\mathbf{SB}_n$ with degree at most $2^k-1$, denoted by $\mathbf{SB}_n^{2^k-1}$, is the ring $<\sigma_1,\sigma_2,\sigma_4,\cdots,\sigma_{2^{k-1}}>$ and isomorphic to $\mathbf{B}_k$.
\end{theo}
\begin{proof}
From Theorem \ref{factor} we know $\mathbf{SB}_n^{2^k-1}$ is contained in $R=<\sigma_1,\sigma_2,\sigma_4,\cdots,\sigma_{2^{k-1}}>$. 
We just check that $R$ is an isomorphism of $\mathbf{B}_k$ since $|\mathbf{B}_k|=|\mathbf{SB}_n^{2^k-1}|$.
Let $$\tau: \mathbf{B}_k\rightarrow R,~f(y_1,y_2,\cdots,y_k)\mapsto f(\sigma_1,\sigma_2,\sigma_4,\cdots,\sigma_{2^{k-1}}).$$
From Theorem \ref{factor} we know $\tau$ is a injection. By Lemma \ref{basic} we know $\sigma_{2^s}^2=\sigma_{2^s}$, and hence $\tau$ is a surjection. Let $f,g\in \mathbf{B}_k$. It's clear that $\tau(f+g)=\tau(f)+\tau(g)$.
Since~$\tau(y_{s})=\sigma_{2^{s-1}}$ for $1\leq s\leq k$, 
we have
$\tau(\Pi_{s=1}^{k}y_{s}^{c_s})=\Pi_{s=1}^{k}\sigma_{2^{s-1}}^{c_s}=\Pi_{s=1}^{k}\tau^{c_s}(y_{s})$ and therefore $\tau(fg)=\tau(f)\tau(g)$.
\end{proof}
\begin{coro}\label{basic_coro:Ring_all}
$\mathbf{SB}_{2^m-1}$ is an isomorphism of $\mathbf{B}_m$.
\end{coro}
The upper degree of the product of two symmetric Boolean functions
is given as below.
\begin{coro}\label{basic_coro}
Let $g,f\in\mathbf{SB}_n$, $\deg(g)\leq 2^{k}-1$ and $\deg(f)\leq
t\cdot2^{k}-1$ ($t\geq 1$). Then $\deg(gf)\leq t\cdot2^{k}-1$.
\end{coro}
\begin{proof}
It holds for $t=1$ since $\mathbf{SB}_n^{2^k-1}$ is a ring by Theorem \ref{basic_coro:Ring}.
Consider the product $\sigma_i\sigma_j$ with $i\leq 2^{k}-1$ and $2^{k}\leq j\leq t\cdot2^{k}-1$ when $t>1$. Let $j=t'\cdot2^{k}+j'$, $1\leq t'\leq t-1$, $j'\leq2^k-1$. From Corollary \ref{basic_coro:1} we have
$\sigma_i\sigma_j=\sigma_i\sigma_{t'\cdot2^{k}+j'}=\sigma_i\sigma_{j'}\sigma_{t'\cdot2^{k}}$, which has degree at most $t'\cdot2^{k}+2^k-1\leq t\cdot2^{k}-1$ since $\sigma_i\sigma_{j'}\in\mathbf{SB}_n^{2^k-1}$.
\end{proof}

From the above corollary we know $\mathbf{SB}_n^{2^k-1}\mathbf{SB}_n^{t\cdot2^k-1}=\mathbf{SB}_n^{t\cdot2^k-1}$ for $t\geq1$.

\section{Fast algebraic attacks on symmetric Boolean functions}\label{sec:FAA}
\noindent In this section, we will first show that fast algebraic attacks on
symmetric Boolean functions work efficiently, and then prove the nonexistence of symmetric FFAI functions.

\begin{theo}\label{deg1_FAA}
Let $f\in\mathbf{SB}_n$, $f(x)=\sigma_{2t+1}+\sum_{i=0}^{2t}\lambda_f (i)\sigma_i$ and $g(x)=\sigma_1+\lambda_f(2t)+1$. Then the multiple $gf$ has degree at most $2t-1$. Moreover, if $\lambda_f(2t)=0$,
then $(\sigma_1+1)f$, if nonzero, has degree $2s+1$ where $s$ is maximum such that $\lambda_f
(2s)=1$; if $\lambda_f (2t)=1$, then $\sigma_1f$, if nonzero, has degree $2s+1$ where $s$ is
maximum such that $\lambda_f (2s)+\lambda_f(2s+1)=1$.
\end{theo}
\begin{proof}
It's trivial for $t=0$. Then assume $t\geq1$.

By Corollary \ref{basic_coro:1}, we have
$\sigma_1\sigma_{2i}=\sigma_{2i+1}$ and therefore $\sigma_1\sigma_{2i+1}=\sigma_{2i+1}$. Hence
\begin{align*}
 h(x)=&(\sigma_1+\lambda_f (2t)+1)f(x)\\
 =&(\sigma_1+\lambda_f (2t)+1)(\sigma_{2t+1}+\lambda_f(2t)\sigma_{2t}+\sum_{i=0}^{2t-1}\lambda_f (i)\sigma_i)\\
 =&\sum_{i=0}^{t-1}[\lambda_f (2t)\lambda_f (2i+1)+\lambda_f (2i)]\sigma_{2i+1}\\
 &+\sum_{i=0}^{t-1}[ (\lambda_f (2t)+1)\lambda_f (2i)]\sigma_{2i},
\end{align*}
showing
$\deg(h)\leq 2t-1.$

If $\lambda_f (2t)=0$, then
$$h(x)=\sum_{i=0}^{t-1}\lambda_f
(2i)\sigma_{2i+1}+\sum_{i=0}^{t-1}\lambda_f (2i)\sigma_{2i}$$ and
therefore $\deg (h)=2s+1$ when $s$ is maximum such that $\lambda_f
(2s)=1$.

If $\lambda_f (2t)=1$, then
$$h(x)=\sum_{i=0}^{t-1}[\lambda_f (2i+1)+\lambda_f
(2i)]\sigma_{2i+1}$$ and therefore $\deg (h)=2s+1$ when $s$ is
maximum such that $\lambda_f(2s)+\lambda_f(2s+1)=1$.
\end{proof}
\begin{rem}
If $\lambda_f(2s)=0$ (resp. $\lambda_f (2s)=\lambda_f (2s+1)$) for any $s$ with $0\leq s\leq t$,
then we have $(\sigma_1+1)f=0$ (resp. $\sigma_1f=0$).
\end{rem}

Theorem \ref{deg1_FAA} gives an affine function $g$
such that $gf$ has degree at most $\deg(f)-2$ for odd $\deg(f)$. In
other words, any symmetric function with odd degree has FAI
strictly smaller than its degree. Although the product $gf$ has odd degree if $gf\neq0$, we cannot apply the theorem recursively to $gf$ to lower the degree of $f$ since $g\cdot gf=g\cdot f$ and $(g+1)\cdot gf=0\cdot f$.

From Theorem \ref{deg1_FAA}, we know that $2^{-t}$ is the probability that $gf=0$ and $2^{-i}$ the probability that $gf$ has degree $\deg(f)-2i$. Consequently, the expectation of the degree of
$gf$ is $\deg(f)-4$ when $\deg(f)$ is large.
\begin{coro}\label{coro:deg1}
The expectation of the degree of the product $gf$ of Theorem \ref{deg1_FAA} is $\deg(f)-4$ when $\deg(f)$ tends
to infinity.
\end{coro}

Now we consider symmetric Boolean functions with degree not equal to
a power of 2.

\begin{theo}\label{deg_FAA}
Let $f\in\mathbf{SB}_n$ and $\deg(f)\geq 2^{k }>1$. If $2^{k }$ does
not divide $\deg(f)$, then there exists a nonconstant function $g$ of degree at most $e$ with $e=\deg(f)~mod~2^{k }$ such that the product $gf$ has degree at
most $\deg(f)-e-1$.
\end{theo}
\begin{proof}
Let $\deg(f)=t\cdot2^{k }+e, t\geq 1, 0<e<2^{k }.$ Let $f
(x)=\sigma_{t\cdot2^{k }+e}+\sum_{i=0}^{t\cdot2^{k }+e-1}\lambda_f
(i)\sigma_i$. By Corollary \ref{basic_coro:1}, we have $\sigma_{t\cdot2^{k
}+i}=\sigma_{t\cdot2^{k }}\sigma_{i}$ for $0\leq i<2^{k }$, and
therefore
$$f (x)=\sigma_{t\cdot2^{k }} (\sigma_{e}+\sum_{i=0}^{e-1}\lambda_f (t\cdot2^{k }+i)\sigma_i)+\sum_{i=0}^{t\cdot2^{k }-1}\lambda_f (i)\sigma_i.$$
Let $g(x)=\sigma_{e}+\sum_{i=0}^{e-1}\lambda_f (t\cdot2^{k
}+i)\sigma_i+1$ and $f^- (x)=\sum_{i=0}^{t\cdot2^{k }-1}\lambda_f
(i)\sigma_i$. Then
$$f (x)=\sigma_{t\cdot2^{k }} (g (x)+1)+f^- (x),$$ and hence $gf=gf^-$. On one hand, the
symmetric function $g$ has degree $e$; on the other hand, by
Corollary \ref{basic_coro}, the function $gf^-$ has degree at most
$t\cdot2^{k }-1=\deg(f)-e-1$.
\end{proof}

The theorem not only proves the existence of the function $g$ but also explicitly identifies several such functions. More exactly, the number of $g$'s is 1 less than the weight
of $\deg(f)$.

Taking $k=\lfloor\log_2\deg(f)\rfloor$, if $\deg(f)\neq 2^k$ then there is a nonconstant function $g$ such that $\deg(g)+\deg(gf)\leq\deg(f)-1$ and therefore the following result is obtained.

\begin{coro}\label{coro:FAI:0}
Let $f\in\mathbf{SB}_n$ and $\deg(f)>1$ is not a power of 2. Then $\mathcal{FAI}(f)\leq \deg(f)-1$.
\end{coro}

Theorem \ref{deg_FAA} and Corollary \ref{coro:FAI:0} show that symmetric functions with degree not equal to a power of 2 do not behave well against fast algebraic attacks.
Then we consider the symmetric functions with degree $2^{\lfloor\log_2n\rfloor}$.
For the case
$n-2^{\lfloor\log_2n\rfloor}$ large,
$2^{\lfloor\log_2n\rfloor}$ is very small compared with $n$ and therefore the symmetric functions with degree $2^{\lfloor\log_2n\rfloor}$ naturally behave badly
against fast algebraic attacks. For $n-2^{\lfloor\log_2n\rfloor}$
not too large, we will show fast algebraic attacks on the symmetric functions with any degree are also very efficient. These imply that almost all symmetric Boolean functions are vulnerable to fast
algebraic attacks.

Now we consider the symmetric functions on $n$ variables, including
the functions of degree equal to a power of 2, for the case
$n-2^{\lfloor\log_2n\rfloor}$ smaller than
$2^{\lfloor\log_2n\rfloor}/2-1$.

\begin{theo}\label{FAA:2}
Let $f\in\mathbf{SB}_n$ and $2^m\leq n<2^{m}+2^{m-1}-1$. Then
$\mathcal{AI}(f)\leq 2^{m-1}-1$ or $\deg(\sigma_ef)=2^{m-1}+e$ with $e=n-2^{m}+1$.
\end{theo}
\begin{proof}
By Theorem \ref{factor}, we have
$$f(x)=\sigma_{2^m}f_{m}
(x)+\sigma_{2^{m-1}}f_{m-1}(x)+f_{m-1}^-(x),$$ where $f_{m}$ is a symmetric
function of degree at most $2^m-1$, and $f_{m-1}$,$f_{m-1}^-$ are of
degree at most $2^{m-1}-1$. Let $g=\sigma_{e}(f_{m-1}+1)$. Since
$n<2^{m}+2^{m-1}-1$, we have $e=n-2^{m}+1<2^{m-1}$ and
therefore $\deg(g)\leq 2^{m-1}-1$ by Corollary \ref{basic_coro}.
By Corollary \ref{basic_coro:2}, we have $\sigma_{e}\sigma_{2^m}=0$ since
$n-2^{m}<e<2^m$. If $g\neq0$, then $gf=gf_{m-1}^-$ which is again of
degree at most $2^{m-1}-1$  by Corollary \ref{basic_coro}. This
means $f$ or $f+1$ admits an annihilator of degree at most
$2^{m-1}-1$, that is, $\mathcal{AI}(f)\leq2^{m-1}-1$. Otherwise
$g=0$, then $\sigma_{e}f_{m-1}=\sigma_{e}$ and hence
$\sigma_{e}f=\sigma_{2^{m-1}+\sigma_{e}}+\sigma_{e}f_{m-1}^-$, which
is of degree $2^{m-1}+e$.
\end{proof}

\begin{rem}
The same proof shows that the theorem applies to $e$ with $n-2^{m}<e<2^{m-1}$, but
$e=n-2^{m}+1$ is minimum.
\end{rem}

Theorem \ref{FAA:2} shows that symmetric functions
on $n$ variables with $n-2^{\lfloor\log_2n\rfloor}$ not large 
are vulnerable to fast algebraic attacks. Especially if $n$ is
close to $2^{\lfloor\log_2n\rfloor}$,
then $e=n-2^{\lfloor\log_2n\rfloor}+1$ is close to 1 and
$d=2^{\lfloor\log_2n\rfloor}/2+e$ is close to ${n}/{2}$, so $e+d$ is close to ${n}/{2}$, and therefore the symmetric functions with AI at least
$2^{\lfloor\log_2n\rfloor}/2$ are very vulnerable to fast
algebraic attacks. For example, any symmetric MAI function $f$
on $2^m$ variables admits the linear
function $\sigma_1$ such
that $\sigma_1f$ has degree $2^{m-1}+1$ while any symmetric function $f$ on
$2^m+1$ variables with MAI $2^{m-1}+1$ or sub-MAI $2^{m-1}$ admits the quadratic
function $\sigma_2$ such that $\sigma_2f$ has degree $2^{m-1}+2$.
They are almost the worst cases against fast algebraic attacks since any function with AI $a$ has FAI at least $a+1$.
Unfortunately, the symmetric MAI functions obtained in \cite{sys_qu2,sys_liu} are in the case $n=2^m$
and the symmetric sub-MAI functions derived in \cite{sys_liao} have $n=2^m+1$.
Moreover, the symmetric MAI functions constructed in \cite[Theorem 2.4]{sys_qu3} have $n\in[2^{m},{5}/{4}\cdot2^{m}]$, and therefore these functions admit $\sigma_e$ with $e\leq n/5$ such that $\sigma_ef$ has degree at most $3n/5$.

Theorem \ref{FAA:2} also gives $g=\sigma_{e}(f_{m-1}+1)\neq0$ and $h=\sigma_{e}(f_{m-1}+1)f_{m-1}^-$ both with degree
at most $2^{m-1}-1$ or $g=\sigma_{e}$ with degree $e$ and $h=\sigma_{2^{m-1}+e}+\sigma_{e}f_{m-1}^-$ with degree $2^{m-1}+e$
such that $gf=h$. The symmetric functions with AI at
most $2^{m-1}-1=2^{\lfloor\log_2n\rfloor}/2-1$ naturally have FAI at
most $2^{m}-2=2^{\lfloor\log_2n\rfloor}-2$. And these symmetric
functions with AI at least $2^{m-1}=2^{\lfloor\log_2n\rfloor}/2$ have FAI smaller than or equal to
$2^{m-1}+2e=2^{m-1}+2(n-2^{m}+1)=2n-3\cdot 2^{m-1}+2=2n-3\cdot2^{\lfloor\log_2n\rfloor}/2+2$.
\begin{coro}\label{coro:FAI}
Let $f\in\mathbf{SB}_n$ and $2^m\leq n<2^{m}+2^{m-1}-1$. Then $\mathcal{FAI}(f)\leq \max\{2^{m}-2,2n-3\cdot 2^{m-1}+2\}.$
\end{coro}

The following theorem proves the nonexistence of symmetric FFAI functions.
\begin{theo}\label{FAA:3}
Let $n\geq 5$ and $f\in\mathbf{SB}_n$. 
Then $\mathcal{FAI}(f)<n$. 
\end{theo}
\begin{proof}
Corollary \ref{coro:FAI:0} has proven the case that $f$ has degree not
equal to $2^{m}$. When $f$ has degree $2^{m}$, we just check the
cases $n=2^{m}$, $2^{m}+1$ or $2^{m}+2$. These cases have been proven
in Corollary \ref{coro:FAI} if
$2n-3\cdot 2^{m-1}+2<n<2^{m}+2^{m-1}-1$, i.e. $m\geq 3$ for
$n=2^m$, $2^{m}+1$ and $m\geq 4$ for $n=2^m+2$. The rest cases
$n=5,6,10$ are confirmed by computing all possible values of FAI for the symmetric functions on $5$, $6$ or $10$ variables.
\end{proof}
%

\section{Relations between algebraic degree and algebraic
immunity of symmetric Boolean functions}\label{sec:Relations}
\noindent  In this section, we will study the relations between algebraic degree and algebraic immunity of
symmetric functions.
It's well known that for any Boolean function $f$ the algebraic immunity is less than or equal to its algebraic degree since $f(f+1)=0$, whereas the relations between algebraic degree and algebraic
immunity can be improved for symmetric functions.
%
\begin{prop}\label{deg_AI1}
Let $f\in\mathbf{SB}_n$. If $f$ has degree not equal to a power of
$2$, then $\mathcal{AI} (f)< 2^{\lfloor\log_2\deg (f)\rfloor}$.
Consequently, we have $\mathcal{AI} (f)\leq 2^{\lfloor\log_2\deg
(f)\rfloor}$ for any $f\in\mathbf{SB}_n$.
\end{prop}
\begin{proof}
Let $k=\lfloor\log_2\deg (f)\rfloor$.
By Theorem \ref{factor} we have
$f=\sigma_{2^k}f_k+f_{k}^-$, where $f_k$ is a symmetric
function of degree $\deg(f)-2^{k}$ and $f_{k}^-$ of degree at most $2^{k}-1$. Let $g=f_k+1$ and $h=gf_{k}^-$. Then $gf=h$, $\deg(g)=\deg(f)-2^{k}\leq 2^k-1$ and $\deg(h)\leq 2^{k}-1$. If $\deg(f)\neq2^k$, then $g\neq0$ and therefore $\mathcal{AI}(f)\leq 2^{k}-1$.
If $\deg
(f)=2^k$, then $\mathcal{AI}(f)\leq \deg (f)=2^{k }$.
\end{proof}
\begin{coro}\label{deg_AI2}
Let $f\in\mathbf{SB}_n$. Then $\deg(f)\geq2^{\lceil\log_2\mathcal{AI}(f)\rceil}
=2^{\lfloor\log_2 (2\mathcal{AI}(f)-1)\rfloor}$.
\end{coro}
\begin{proof}
We only check the case $\mathcal{AI} (f)>1$. Let $a=\mathcal{AI}(f)$ and
$d=\deg (f)$. Proposition \ref{deg_AI1} shows that $\log_2a\leq
{\lfloor\log_2d\rfloor}$, i.e. $\lceil\log_2a\rceil\leq
{\lfloor\log_2d\rfloor}$. Hence $d \geq
2^{\lceil\log_2a\rceil}=2^{\lfloor\log_2 (2a-1)\rfloor}$.
\end{proof}

Siegenthaler's inequality\cite{Siegenthaler} states that any $m$-th order correlation-immune
function has degree at most $n-m$ and any
$m$-resilient function ($0\leq m<n-1$) has degree at most $n-m-1$.
Therefore the order of correlation-immune (resp. resiliency) of any symmetric Boolean function with AI equal to $a$ $(a>1)$ is smaller
than or equal to $n-2^{{\lfloor\log_2 (2a-1)\rfloor}}$ (resp. $n-2^{{\lfloor\log_2 (2a-1)\rfloor}}-1$).

Now we consider the lower bound of algebraic degree for symmetric MAI functions.
\begin{coro}\label{deg_MAI}
Let $f\in\mathbf{SB}_n$ and $\mathcal{AI}
(f)=\lceil\frac{n}{2}\rceil$. Then $\deg (f)\geq 2^{\lfloor\log_2(n-1)\rfloor}$.
\end{coro}
\begin{proof}
Since $\lceil\frac{n}{2}\rceil\geq\frac{n}{2}$, by Corollary
\ref{deg_AI2}, we have $\deg (f)\geq 2^{\lfloor\log_2
(n-1)\rfloor}$.
\end{proof}

For every $n$, there exist symmetric MAI functions on $n$
variables of degree $2^{\lfloor\log_2n\rfloor}$. For
example, the majority function $f$
achieves MAI and also has degree
$2^{\lfloor\log_2n\rfloor}$ \cite{Dalai}. When $n=
2^m$, the function $\sigma_{2^{m-1}}$ achieves MAI \cite{Braeken}.
In addition, all the symmetric MAI functions on $2^m$ variables
were obtained in \cite{sys_qu2,sys_liu} and were proven
having algebraic degree $2^{m-1}$ or $2^m$.
Notice that $2^{\lfloor\log_2(n-1)\rfloor}=2^{\lfloor\log_2n\rfloor}$ if $n\neq 2^m$, and
$2^{\lfloor\log_2(n-1)\rfloor}=2^{m-1}$ if $n=2^m$.
Therefore the bound of Corollary \ref{deg_MAI} is tight.



\begin{table}[!h]
\tabcolsep 0pt \caption{The upper algebraic immunity
of symmetric functions with designated degree}\label{table:1} \vspace*{-8pt}
\begin{center}
\def\temptablewidth{.96\textwidth}
{\rule{\temptablewidth}{1pt}}
\begin{tabular*}{\temptablewidth}{@{\extracolsep{\fill}}@{~}c@{~}|c|cccccccc@{~}}
$\deg$       & $d$ &1 & 2--3 & 4--7 & 8--15& 16--31& 32--63&64--127&128--255\\\hline
Upper AI & $2^{\lfloor\log_2d\rfloor}$ &1 & 2 & 4   & 8  & 16 & 32 & 64 & 128 \\
\end{tabular*}
       {\rule{\temptablewidth}{1pt}}
\end{center}
\end{table}

\begin{table}[!h]
\tabcolsep 0pt \caption{The lower degree 
of symmetric functions with designated algebraic immunity}\label{table:2} \vspace*{-8pt}
\begin{center}
\def\temptablewidth{.96\textwidth}
{\rule{\temptablewidth}{1pt}}
\begin{tabular*}{\temptablewidth}{@{\extracolsep{\fill}}@{~}c@{~}|c|cccccccc@{~}}
AI        & $a$ &1 & 2 & 3--4 & 5--8& 9--16& 17--32&33--64&65--128\\\hline
Lower $\deg$ & $2^{\lceil\log_2a\rceil}$ &1 & 2 & 4   & 8  & 16 & 32 & 64 & 128 \\
\end{tabular*}
%
       {\rule{\temptablewidth}{1pt}}
\end{center}
\end{table}

Notice that the relation between algebraic degree and algebraic immunity of symmetric
Boolean functions doesn't relate to the number of variables.
Therefore the bounds listed in Table \ref{table:1} and Table \ref{table:2} are true for any reasonable $n$.
We leave a open problem whether the bound is tight when AI isn't MAI.

\section{Conclusion}\label{sec:Con}
Symmetric Boolean functions, which can be considered as compositions of Boolean functions and elementary symmetric functions with power-of-2 degree, behave badly against fast algebraic attacks, so these functions are unfit to
be used in stream ciphers. 
In other words, if symmetric functions are used in the design of ciphers, fast algebraic immunity should never be ignored, and the number $n$ of variables had better be neither equal to nor a little more than $2^m$. $n$ approximating $3\cdot2^{m-1}$ seems to be a good choice
but it still need further study.


\end{document}